\documentclass[letterpaper,11pt]{report}

\usepackage{fullpage}
\usepackage{verbatim}
\usepackage{cite}
\usepackage{setspace}
\usepackage{fancyhdr}
\usepackage[a4paper]{geometry}

\usepackage[small]{caption}
\usepackage{graphics}
\usepackage{color}
\usepackage{upquote}
\usepackage{algorithm2e}
\usepackage{hyperref}
\usepackage{multirow} % http://ctan.org/pkg/multirow
\usepackage{rotating}
\usepackage{fixltx2e}
\usepackage{mathtools}
\usepackage{float}
\usepackage{fancybox}
\usepackage[font={small,it}]{caption}

\usepackage{hyperref}

\def\title{Mytitle}

\begin{document}

\def\addrone{Your address}
\def\addrtwo{Your city}

\def\degree{M.Tech. in Computer Science with Specialization in Information Security}

\def\submissiondate{November 20, 2013}

\def\supervisorone{Dr. Ponnurangam Kumaraguru}

\def\supervisortwo{Dr. Vinayak Naik}

\def\supervisorthree{Dr. Muttukrishnan Rajarajan}

\ %def\supervisorfive{YYY YYYY}

\thispagestyle{empty}

\begin{center}

{\LARGE \bf {OCEAN: Open-source Collation of eGovernment data And Networks }

 }
 \vspace{.3in}

 {\Large{Student Name: Srishti Gupta}} \\
 \vspace{.1in}
 IIIT-D-MTech-CS-IS-13-MT11012 \\

 November 20, 2013 \\

    \vspace{.35in}

  \vspace{.25in}

{Indraprastha Institute of Information Technology\\
New Delhi}

\vspace{.35in}  {\underline{Thesis Committee} \\ \supervisorone                ~(Chair) \\ \supervisortwo \\ \supervisorthree }\\ \vspace{.35in}

 {Submitted in partial fulfillment of the requirements \\for the Degree of M.Tech. in Computer Science, \\ with specialization in Information Security}

\vspace{.2in}

\copyright 2013  IIIT-D-MTech-CS-IS-13-MT11012 \\ All rights reserved \\
\vspace{.8in}

\end{center}

%This research was partially funded by XXXX YYYY.

\newpage

\pagestyle{empty}
\vspace*{7.1in}

\newpage

\begin{center}
\section*{Certificate}\label{section:certificate}
\end{center}
%\vspace{3in}
This is to certify that the thesis titled \textbf{``OCEAN: Open-source Collation of eGovernment data And Networks"} submitted by \textbf{Srishti Gupta} for the partial fulfillment of the requirements for the degree of \emph{Master of Technology} in \emph{Computer Science \& Engineering} is a record of the bonafide work carried out by her under my guidance and supervision in the Security and Privacy group at Indraprastha Institute of Information Technology, Delhi. This work has not been submitted anywhere else for the reward of any other degree. \\ \vspace{0.5in}

\textbf{Professor Ponnurangam Kumaraguru}\\
\textbf{Indraprastha Institute of Information Technology, New Delhi}
%\doublespacing

\begin{abstract}

The awareness and sense of privacy has increased in the minds of people over the past few years. Earlier, people were not very restrictive in sharing their personal information, but now they are more cautious in sharing it with strangers, either in person or online. With such privacy expectations and attitude of people, it is difficult to embrace the fact that a lot of information is publicly available on the web. Information portals in the form of the e-governance websites run by Delhi Government in India provide access to such PII without any anonymization. Several databases e.g., Voterrolls, Driving Licence number, MTNL phone directory, PAN card serve as repositories of personal information of Delhi residents. This large amount of available personal information can be exploited due to the absence of proper written law on privacy in India. PII can also be collected from various social networking sites like Facebook, Twitter, GooglePlus etc. where the users share some information about them. Since users themselves put this information, it may not be considered as a privacy breach, but if the information is aggregated, it may give out much more information resulting in a bigger threat. For e.g., data from social networks and open government databases can be combined together to connect an online identity to a real world identity. Even though the awareness about privacy has increased, the threats possible due to the availability of this large amount of personal data is still unknown. To bring such issues to public notice, we developed Open-source Collation of eGovernment data And Networks (OCEAN), \footnote{http://precog.iiitd.edu.in/research/ocean.} a system where the user enters little information (e.g. Name) about a person and gets large amount of personal information about him / her like name, age, address, date of birth, mother's name, father's name, voter ID, driving licence number, PAN. On aggregation of information within the Voter ID database, OCEAN \footnote{OCEAN: Best poster award, IIT Kanpur Symposium on Cyber Security, 2013} creates a \emph{family tree} of the user giving out the details of his / her family members as well. We also calculated a privacy score, which calculates the risk associated with that individual in terms of how much PII of that person is revealed from open government data sources. 1,693 users had the highest privacy score making them the most vulnerable to risks. Using OCEAN, \footnote {OCEAN: Work covered in national newspaper of Delhi, Hindustan in April 2013} we could collect 8,195,053 Voterrolls; 2,24,982 Driving licence; 53,419 PAN card numbers; 1,557,715 Twitter; 3,377,102 Facebook; 29,393 Foursquare; 1,86,798 LinkedIn and 28,900 GooglePlus records. There exist several websites like Yasni, \footnote{http://www.yasni.com/.} PeekYou, \footnote{http://www.peekyou.com/india.} Pipl \footnote{https://pipl.com/.} which help in searching a person on the Internet but are not focused for people living in Delhi. We performed a user evaluation of OCEAN \footnote{OCEAN: Work highlighted on IIIT - Delhi website, \emph{Research} section} in a survey study to evaluate the usability, effectiveness and impact of OCEAN \footnote{OCEAN: Accepted poster at IBM I-care, 2012} and showed that users like and find it convenient to use it in real-world. We received 661 total hits (657 unique visitors) from the day we released the system, January 21, 2013, until October 10, 2013. To the best of our knowledge, this is the first real world deployed tool which provides personal information about residents of Delhi to everyone free of cost.

\end{abstract}

\newpage
\pagestyle{empty}

\newpage

\section*{Acknowledgments}\label{section:acknowledgments}
\pagestyle{plain}
\pagenumbering{roman}

I would like to express my sincere gratitude to my advisor \emph{Dr. PK} for his vision and guidance throughout the project. This thesis would not have been materialized without your support. Your faith and confidence in me helped in exploring all the dimensions of this project. I would like to thank my esteemed committee members \emph{Dr. Vinayak Naik} and \emph{Dr. Muttukrishnan Rajarajan} for agreeing to evaluate my thesis work.
\newline
I thank all the members of Precog research group at IIIT- Delhi for their valuable feedback and suggestions, especially \emph{Niharika Scahdeva} for shepherding me and spending her valuable time to come up with this thesis. I would also like to thank Mayank Gupta (B.Tech, DCE) who helped in developing some modules of OCEAN.
\newline
Last but not the least, I would like to thank all my family members and friends who encouraged and kept me motivated throughout the project.

\newpage

\tableofcontents
\listoffigures
\listoftables

\newpage

\newpage

\newpage
\mbox{}

%\doublespacing
\pagenumbering{arabic}
\chapter{Research Motivation and Aim}\label{chapter:Research Motivation and Aim}
\begin{figure*}[!htp]
\centering
%\vspace{-2.0cm}
\includegraphics[scale=0.45, angle=0]{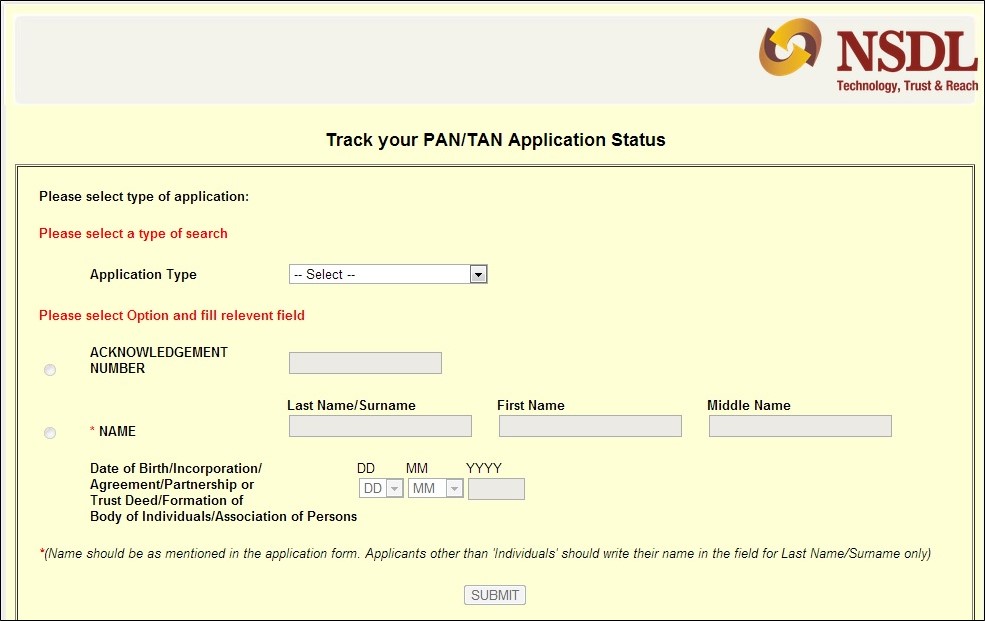}
%\vspace{0.5 cm}
\caption{Public interface of PAN card database of Income Tax department of Delhi. The input is first name, last name and DOB of the person and the output is full name and PAN number.}\label{fig:pan}
\vspace{0.5cm}
\end{figure*}

The Government of India is encouraging all states of the union of India to move to e-governance model and distribute information digitally through websites. \footnote{http://planningcommission.nic.in/plans/planrel/fiveyr/10th/volume2/10th\_vol2.pdf} In it's tenth five year plan, it announced that it should make government processes to be a `SMART'(Simple, Moral, Accountable, Responsible and Transparent) governance. This had led to several e-governance initiatives by many states and resulted in publicly accessible databases. A lot of states in India are storing information about its people digitally i.e., in the form of databases. These data repositories are created in order to improve data availability, make application processes easier, making data available for quicker responses to RTI requests etc. As a case study, we have focussed on the databases maintained by Delhi government only. Government of Delhi has implemented most of the aforementioned e-governance guidelines and created databases for critical IDs like Driving Licence, Voter ID, PAN card, MTNL, Income tax e-filing where the user can query the system and obtain his / her data. All the above databases though are designed to speed up the exchange of information done among various government departments and for easy dispersal of information to the general population, they pose a threat to the privacy of any individual whose data can be obtained by querying the data-stores through their public interfaces. Also, often data obtained from one such source can be used as the input to another source and extract more data. The potential exploitation of individual's privacy is a major concern. A threat model describing threats, security issues and vulnerabilities from these government portals is described in the Section 3.
\newline
The privacy perceptions of Indian citizens have changed in the recent years \cite{kumaraguru:privacy-in-india:-attitud:2012:yuqfj}. They demand for protection of the data they share digitally.  A recent report identifies a number of examples that may be considered PII \cite{mccallisterguide} including: Name (full name, maiden name, mother’s maiden name), personal identiﬁcation number (e.g., Social Security Number), address (street or email address), telephone numbers, or personal characteristics (such as photographic images especially of face or other distinguishing characteristic, X-rays, fingerprints, or other biometrics). The Information Technology Act 2000 of Indian Parliament, \footnote{http://deity.gov.in/sites/upload\_files/dit/files/GSR313E\_10511.pdf} considers any information that directly / indirectly identifies a person as personal information. The offline world of a user, which is characterized by personal attributes like name, address, Date of Birth (DOB), age, father's name / mother's name can be considered as PII since it helps in uniquely identifying a person. This personal information is stored digitally by the Delhi government in the form of several databases like Voter ID, Driving Licence, MTNL phone directory, PAN card number. Figure \ref{fig:pan} shows the public interface of PAN card database of Income Tax department of Delhi.
Even though different government departments maintain individual databases, the data can be aggregated and misused, if needed. However, PII is sensitive, appropriate precautions are required to protect PII, as the unauthorized release or abuse of PII could result in potentially grave repercussions for the individual who’s PII has been compromised. For example, the information can be used to create fake documents, open fake bank accounts, procure phone connection or credit card by pretending to be someone else. PAN card can be used to link a person's credit card, banking, income tax returns, house and other cash or immovable assets data.
\newline
The online world characterized by attributes like name, location, profile images, profile url etc. is shared by the user on various social networking sites like Facebook, Twitter, GooglePlus etc. The profile image which is publicly available on these sites can be morphed and can be used with bad intentions. Since users themselves share this information, it cannot be said as a privacy loss, but if all these data can be combined together, it will give out much more information than desired by the users. Several cases \footnote{http://www.hindustantimes.com/India-news/Gurgaon/Identity-theft-cases-on-the-rise/Article1-931638.aspx} exist to show the rise of identity thefts using the personal information obtained from these social networks. For e.g., the malicious users can access accounts of individuals on various social networking sites such as Facebook, Twitter etc. and retrieve their photos and other information and use the same for making fake driving licences, applying for telephone connections, opening bank accounts and making PAN cards and credit cards.
\section{Vulnerabilities in Open Government Data}
The various countries in the world, including US, \footnote{http://www.archives.gov/open/available-datasets.html} UK \footnote{http://data.gov.uk/data/search} and India \footnote{http://data.gov.in/} have started open government initiatives to make some information accessible to people. In this thesis work, we have focussed on the publicly available datasets from government of India and discuss the vulnerabilities existing from these sources in this section. The information obtained from one open government source can be used as an input to other open government source and can be used to extract highly sensitive information. The findings are discussed below.
\subsection{Income Tax Returns}
\begin{figure*}[!htp]
\centering
%\vspace{-2.0cm}
\includegraphics[scale=0.45, angle=0]{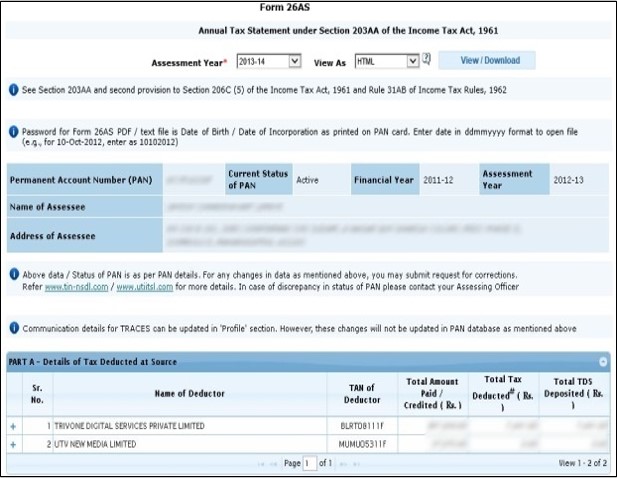}
%\vspace{0.5 cm}
\caption{Screenshot showing the tax statement of a random person on Income tax e-filing portal.}\label{fig:tds}
\vspace{0.5cm}
\end{figure*}

An open government initiative by Income Tax Department, Government of India, allows individuals to pay their tax online. \footnote{https://incometaxindiaefiling.gov.in/} The registration page needs details like PAN number, fullname, DOB, email-ID and phone number. The first three set of information is available from the Voter ID and PAN card databases of the Government of Delhi. The last two can be entered as dummy / fake values. Once the user is registered, he / she is asked to fill a registration form specifying the address and setting a password for himself / herself. The activation link sent to the fake e-mail account verifies and registers the user on the portal. Once the user gets registered, we can view the Form 26AS which shows the tax statement (tax credit) of the individual. The form shows the total amount of tax paid by the individual in the chosen financial year (for e.g., 2013-2014 in this case). Figure \ref{fig:tds} shows the tax statement of a random person as shown in this article. \footnote{http://www.techtree.com/content/features/4227/guide-how-file-income-tax-returns-online-2013-edition.html}
(\emph{Note: The image is taken from the article. \footnote{http://www.techtree.com/content/features/4227/guide-how-file-income-tax-returns-online-2013-edition.html}})
\subsection{Online Voter ID card}
\begin{figure*}[!htp]
\centering
%\vspace{-2.0cm}
\includegraphics[scale=0.50, angle=0]{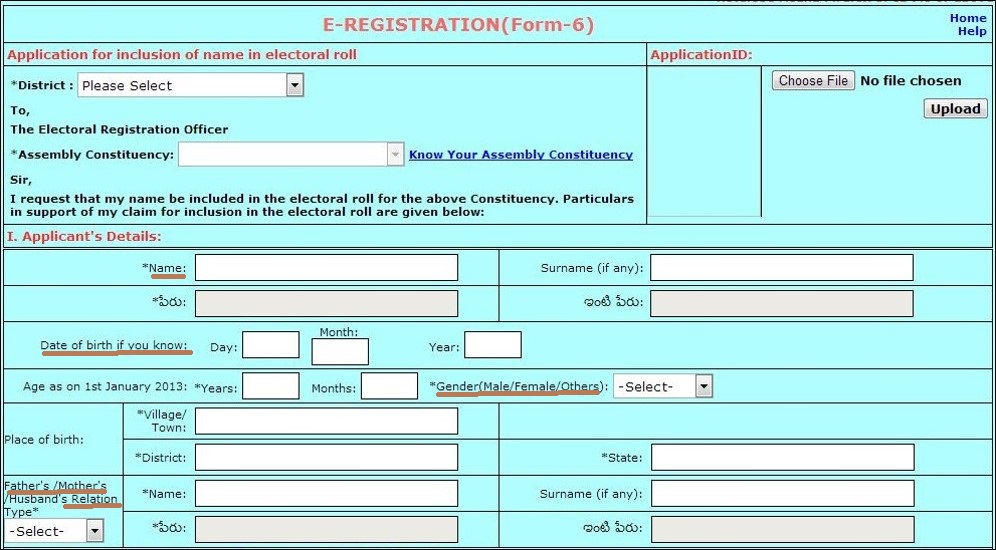}
%\vspace{0.5 cm}
\caption{Screenshot showing the application form of a random user picked from OCEAN database on online Voter ID portal.}\label{fig:voter-2}
\vspace{0.5cm}
\end{figure*}

Today, Voter ID card is a must in India. Every law abiding citizen is bound to have this identity card. The Government of India has set up a portal / service which allows people to get a Voter ID card online. The ease with which multiple fake Voter ID cards can be generated is of great concern. The registration page for the portal \footnote{http://eci-citizenservices.nic.in/} requires a mobile number and an e-mail account (which can be faked as described in the previous section). After the registration, the user is asked to fill a form entering his / her name, gender, town / district. All this information is available from the Voter ID database of the Government of Delhi and can be used directly. Thereafter, the user is required to fill another form giving a detailed set of personal information like name, DOB, age, address, gender, family details as shown in the Figure \ref{fig:voter-2}. The above-mentioned information is available in the Voter ID database of the Delhi government. On completion of this form, an application ID will be issued to the user which can be used by him / her to trace the status of the online voter card. The ease of getting into such open government systems with all the personal information beforehand indicates the alarming situation and a pertaining need to take immediate actions to help resolve these issues.
\newline
Other open government initiatives like BSES Electricity Bill Payment portal \footnote{http://www.bsesdelhi.com/bsesdelhi/caVerification4Pay.do} can be exploited which takes CA number as the input and name, address, load value, circle, district, current demand details as the output. The incremental nature of the 9-digit CA number makes it vulnerable to attacks. The examples demonstrate the privacy breach of the citizens in Delhi in revealing their personal information on the web by various open government initiatives.
\section{Research Aim}
To educate people on the privacy issues and risks arising from open government data, we made a first attempt to deploy an integrated system, OCEAN: Open-source Collation of eGovernment data And Networks, where the user can enter little amount of information (e.g. name) about a person and get large amount of information about him / her: Voter ID, name, address, age, DOB, father`s name / mother`s name, driving licence number, PAN card number and account information on social networking sites viz., Twitter, Facebook, Foursquare, LinkedIn, GooglePlus.
\newline
OCEAN also aims to demonstrate that aggregation of the available information yields out much more information about Delhi residents. It creates a \emph{family tree} for a random user by aggregating the results within the Voter ID database which helps in finding the details of other family members. So, the aim of OCEAN is
\begin{itemize}
\item[$\bullet$]
To develop a technology to demonstrate the public availability of large amount of personal information on the web by open government data sources.
\item[$\bullet$]
To highlight the privacy issues arising out on the aggregation of this personal information on Delhi citizens.
\item[$\bullet$]
To develop an empirical understanding on awareness, experience and expectations of Delhi citizens in using the open government data sources.
\end{itemize}

\chapter{Related Work and Research Contributions}\label{chapter:Related Work and Research Contributions}
\section{Related Work}
Personally Identifiable Information (PII) is one of the most central concepts in information privacy regulation \cite{schwartz2011pii}. The scope of privacy laws typically turns on when PII is involved. At the same time, there is no uniform definition of PII in the data protection law in India. Moreover, computer science has shown that in many circumstances non-PII can be linked to individuals, and that de-identified data can be re-identified. This section gives an overview of the related research showing the privacy risks and attack on the information available from open government data and social networks.
\subsection{Privacy on the web / Open Government Data}
Digital identities and profiles are precious assets. On one hand they enable users to engage in transactions and interactions on the Internet. On the other hand, abuses and leakages of this information can violate the privacy of their owners, sometimes with serious consequences \cite{mont2003towards}. A recent article \footnote{http://moneylife.in/article/sachin-dhoni-shah-rukh-and-salmans-i-t-accounts-too-hacked-by-ca-student/34620.html} shows that the IT-accounts of famous celebrities in India could be accessed by a common man. The details were fetched from the e-filing Income tax portal run by government of Delhi by exploiting the vulnerabilities in the system. Several articles \footnote{http://articles.timesofindia.indiatimes.com/2012-05-25/nagpur/31850566\_1\_college-website-engineering-college-college-authorities} \footnote{http://articles.timesofindia.indiatimes.com/2012-08-27/noida/33424309\_1\_dubai-identity-theft-residential-visa} demonstrate the cases of identity thefts where people impersonate an individual with access to his / her personal information in addition to some fake information. Celnet report \cite{godse2010rise} shows that E-transactions account for 30\% of the total transactions and 75\% of the total payment value is now made electronically. With the penetration of e-governance and banking applications, the security of the information of the consumers involved in the transactions is important. Confidential information (including personal data, financial details, business data) needs to be disclosed in order to enable these interactions. The information might disclose personal data to third parties (such as suppliers, information providers, government and financial institutions, etc.) which lead to privacy loss for the individual. It has been observed that user identity may be exploited in attacks on the privacy of users in web search query logs \cite{jones2007know}. It presents a new attack in which a real-world acquaintance of a user attempts to identify that user in a large query log, using personal information. There have been several articles discussing the potential privacy loss due to Open Government data. The article \footnote{http://www.theguardian.com/society/2012/jul/10/open-data-force-for-good-risks} shows that preventing disclosure of personal information, as obtained from open data remains a question. Another article, \footnote{http://manypossibilities.net/2013/06/the-open-data-cart-and-twin-horses-of-accountability-and-innovation/} talks about `Mosaic Effect', which occurs when the information in an individual dataset, in isolation, may not pose a risk of identifying an individual (or threatening some other important interest such as security), but when combined with other available information, could pose a privacy risk. \cite{nashash2013education} shows the risks that private information may somehow be collected and published online, and that in case databases of different agencies where combined, the resulting combination of information might negatively impact them. The availability of personal information in government records is exploited for commercial and sometimes criminal purposes. Barber et al. \cite{barber2006personal} shows that commercial data aggregators routinely mine government records to gather information about individuals in the United States.
\subsection{Privacy in Online Social Networks}
India has experienced enormous development in information technology \cite{godse2010rise}. Over the last decade, Online Social Media has evolved which is used for maintaining social connections, thereby introducing online social networks. According to Nielsen's Social Media Report, users continue to spend more time on social networks than on any other kind of websites on the Internet \cite{Nielsen}. These social networks describe the connections between a user and his / her friends. With the advent of social networks, the Internet community has experienced a revolution in its communication habits. These networks allow people to publish details about themselves. Some information in these networks is private and with some attacks / learning algorithms, undisclosed information might be revealed. For example, Krishnamurthy et.al \cite{bala} showed that indirect leakage of PII via OSN identiﬁers to third-party aggregation servers is happening on these networks. Researchers have shown that BBM pins \footnote{http://precog.iiitd.edu.in/labs/bbmpinleak/} could be extracted from the tweets / posts of users from Twitter and Facebook respectively which could be used in spamming these users. The content generated by people in the social network graph of the person can be used to identify some information about him / her. Zheleva et.al \cite{zheleva2009join} showed that an adversary / malicious user can exploit an online social network with a mixture of public and private user proﬁles to predict the private attributes of users. There has been research done to show that the information about an individual can be inferred from his / her friends in the social network. Jurgens et al. \cite{jurgens2013s} showed that location of the individual could be inferred using the geographic distribution of his friends on Twitter. On the other hand, content generated by one person can be used to gain some information about his / her friends online. Mislove et al. \cite{mislove2010you} showed that given a set of attributes for some section of people, it was possible to infer the attributes of other users connected in the network.
Since large amount of data is flowing across the social networks in digital form, it is impertinent to keep the data secure so that it cannot be used for unlawful activities.
\newline
\newline
As described in this section, OCEAN shows that information aggregation gives us more information; aggregation within open government databases allowed us to create a family tree and combining some profiles from Facebook, Twitter and Foursquare gave us additional information, like location, which is available on Twitter but not on Facebook.
\subsection{Systems built with Open Government Data}
Since lot of datasets and reports are published by the government, a lot of work is being done using this open data. Researchers are working on developing a web portal which will enable users to explore and analyze the open data available in India \cite{sandesh}. There have been several systems / portals developed from the open government data sources as reported in the study by Wright et al. \cite{wright2010open}.
\begin{itemize}
\item[$\bullet$ ] \textbf{IndianKanoon \footnote{http://www.indiankanoon.org/}:}
It is a legal search engine which indexes judgements and statutes of the Supreme Court of India and several High Courts. It also integrates reports of the Law Commission, open access law journals and other online legal repositories.
 \item[$\bullet$ ] \textbf{OpenCivic.in \footnote{http://www.opencivic.in/}:}
The government websites including state assembly elections and profiles of MP's in Maharashtra were scraped to get an idea about civic participation. It then provides an API which provides data in machine readable from which helps developers to create visualizations.
\end{itemize}
There have been several applications developed from the open government data in USA \footnote{http://www.data.gov/developer-apps-showcase} and UK.\footnote{http://data.gov.uk/apps}
\begin{itemize}
\item[$\bullet$ ] \textbf{ABQ Ride \footnote{http://www.cabq.gov/abq-apps/city-apps-listing/abq-ride}:}
The application shows the real-time locations of city buses, within a minute's accuracy in the USA. It can also find bus schedules and fares for other public transportation.
\item[$\bullet$ ] \textbf{Illustreets \footnote{http://data.gov.uk/apps/illustreets}:}
This application puts deprivation, crime, education, transport, environment, and census data on an interactive, searchable map, which helps in comparing between locations on the fly. This application is developed for England, UK.
\end{itemize}

\section{Research Contributions}
\begin{itemize}
\item[$\bullet$]
To the best of our knowledge, this is the first deployed system which shows the aggregated personal information about the residents of Delhi. Few identity search systems, e.g. Yasni, Pipl and PeekYou which uses multiple approaches to search for a user on the Internet including social networks are built to extract maximum possible information from the web about a person, however the result set returned for each user query are large, making the system non-usable. We conducted a survey study to calculate the standard System Usability Score (SUS) to measure the effectiveness, efficiency and satisfaction of the users in using OCEAN.
\item[$\bullet$]\textbf{Privacy Score:}
This score tells the risk associated with the person on the leakage of personally identifiable information about him / her from the open government databases.
\item[$\bullet$]
We did a threat modelling on the various open government databases and calculated a DREAD (Damage, Reproducibility, Exploitability, Affected users, Discoverability) score \footnote{http://en.wikipedia.org/wiki/DREAD:\_Risk\_assessment\_model} to estimate the risks and threats arising due to the availability of personal information from these sources.
\item[$\bullet$]
We developed an empirical understanding of privacy perceptions, awareness of the people about the open government data and the expectations of the users from the government in using these data sources.
\end{itemize}

\chapter{Proposed Methodology}
In this section, we discuss the methodology approach used to develop the system. The system was developed in four phases:
\section{Identification of Available Online Open Government Sources}
 The first phase involved identifying all the databases that hold public information readily available on the Internet. Several databases of Government of Delhi which were looked at are:
 \begin{itemize}
 \item[$\bullet$ ] \textbf{Driving Licence database of the Government of Delhi \footnote{https://www.dimtspay.in/dldetail/default.aspx}:}
 The driving licence in India are issued by individual states which permits a person to drive in the country. One must be 18 years or above to receive this license.
 The input required was a valid driving licence number of the form DL-XXYYYYAAAAAAA where DL denoted the state of Delhi, XX a Location in Delhi, YYYY the year of issue of the license, AAAAAAA unique for the subject in question. Hence, using a random generated driving licence number we were able to find various details for a single user like name, address, father’s /mother's / husband's name, date of birth, validity period, and vehicle category. Repeating the process for different randomly generated licence numbers we were able to find details of many subjects. The driving licence database does not have any security features to limit the number of queries that we could ask the database.
 \item[$\bullet$ ] \textbf{Voter ID database of Election Commissioner of Delhi \footnote{http://ceodelhi.gov.in/OnlineErms/ElectorSearch.aspx}:}
     Voter ID is a unique identity given to a user which allows him / her to vote or receive a ballot for an election. One must be 18 years or above to receive this privilege. This database requires the name of the subject along with the constituency in which the subject resides. According to the chief electoral website of Government of Delhi, the state is divided into 70 constituencies.\footnote{http://www.ceodelhi.gov.in/AccemblyConstituenty.aspx.}. To find the constituency of a particular user, PHP functions were used to extract the constituency name from the complete address using simple string compare functions. The data that can be extracted from this database is name, father’s name, date of birth, gender, ID card number of the card holder. This database also did not have any security feature to prevent us from querying it multiple times.
 \item[$\bullet$ ] \textbf{PAN card database of Income Tax department of Delhi \footnote{https://tin.tin.nsdl.com/tan/servlet/PanStatusTrack}:}
 PAN number is a unique alphanumeric combination issued to all juristic entities identifiable under the Indian Income Tax Act 1961. This number is mandatory for making financial transactions. One need to be 18 years or above to get this number. This database requires the name of the person and DOB as the input. These details could be obtained from the Driving Licence database mentioned above. The data that can be extracted from this database is name, PAN number of the card holder. This database too did not have any security feature to prevent us from querying it multiple times.
  \item[$\bullet$ ] \textbf{Phone number database of MTNL department of Delhi \footnote{http://phonebook.bol.net.in/}:}
  This database requires only the name of the person as the input. The data that can be extracted from this database is name, address and phone number. Since the owning authorities changed the policy recently, this database is not available for public on OCEAN.
 \end{itemize}
 For obtaining the online attributes related to an individual, we collected data from the 5 popular social networks  viz., Twitter, Facebook, Foursquare, LinkedIn, and Google Plus using the APIs provided by these networks.
Table 3.1 shows the information obtained from various open government data sources and social networks.
\begin{table*}[!htp]
\caption{Information extracted from various open government data sources.}\label{tab:metadata}
\begin{center}
\begin{tabular}{|p{3cm}|p{4cm}|p{8cm}|} \hline
\textbf{Database} & \textbf{Input Data} & \textbf{Data Retrieved} \\ \hline
\small Driving Licence & \small Driving Licence number & \small Name, Address, Father's name, DOB, Validity period, license type / vehicle category.\\ \hline
\small Voter ID & \small Name, constituency & \small Voter ID, Name, Address, Father's / Mother's / Husband's name, Age, Gender.\\ \hline
\small PAN Number & \small Name, DOB & \small PAN Number, Name.\\ \hline
\small Phone number & \small Name & \small Name, Address, Phone number.\\ \hline
\small Facebook & \small Name, access token & \small Full name, Facebook ID, Gender, Username, Profile Image, Profile URL.\\ \hline
\small Twitter & \small Name, oauth token & \small Full name, Twitter ID, Screen name, Friends count, Location, Following count, Followers count, Profile image, Profile URL.\\ \hline
\small Foursquare & \small Name, oauth token & \small Foursquare ID, Full name, gender, City, Facebook / Twitter contact, Friend count, Biography, Badge count, Mayorship count, Check-in count, Following count, Profile image, Profile URL.\\ \hline
\small LinkedIn & \small Name, oauth token & \small LinkedIn ID, Full name, Location, Headline, Profile image, Profile URL.\\ \hline
\small Google Plus & \small Name, key& \small GooglePlus ID, Full name, Gender, Tag line, About me, relationship status, Location, Places Lived, Organization, Birthday, E-mail, Language, Profile image, profile URL .\\ \hline
\end{tabular}
\end{center}
\end{table*}
\section{Threat Modelling}
\begin{figure*}[!htp]
\centering
%\vspace{-1.0cm}
\includegraphics[scale=0.40, angle=0]{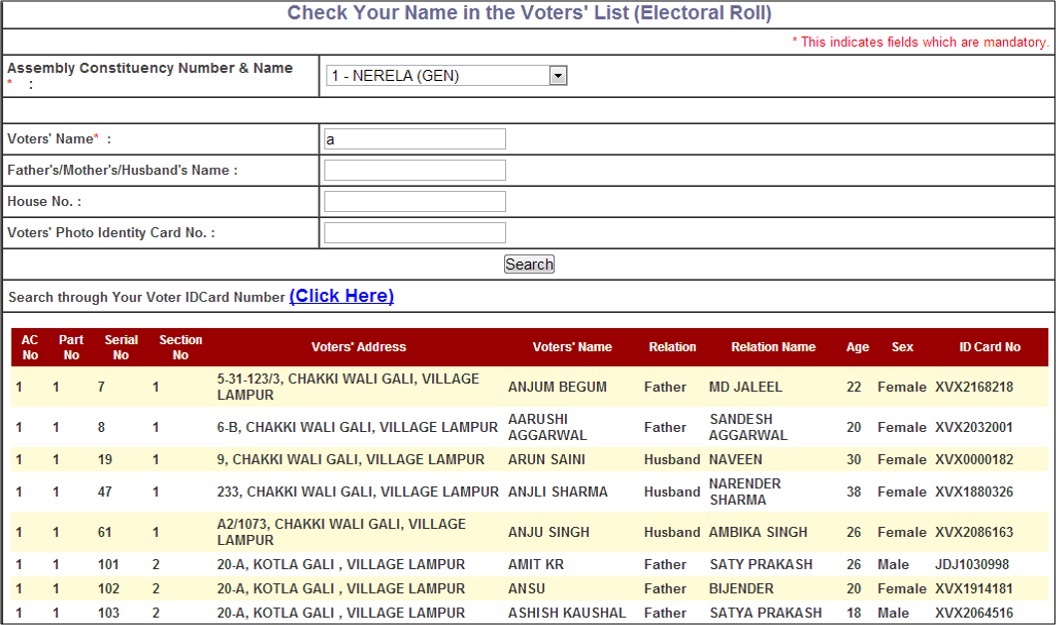}
\caption{Screenshot of Voter ID database showcasing information disclosure. The input for name is given as `a' and output gives name of all the people having names starting with `a'.}\label{fig:motivation}
\vspace{0.5cm}
\end{figure*}

After identifying the open government data sources, we did a threat modelling to identify the risks and threats arising from these repositories. A threat model \footnote{http://en.wikipedia.org/wiki/Threat\_model} is used to identify holes in any software application / system which can be exploited from an adversarial point of view. In this section, we will describe the threat model for the open government databases of government of Delhi. The various public interfaces owned by the Delhi government are at risk since they do not contain well defined privacy policies. The ease with which the information is accessible adds on to the risk level possessed by these open government sources. For e.g., Figure \ref{fig:motivation} shows the website for Voter ID database of Election Commissioner of Delhi which shows the privacy breach / information disclosure by these websites. Entering only an alphabet `a' gives the details of all the individuals whose name starts from `a'. Such an attack can be replicated by a malicious user very easily.
For classifying the security threats of these open government databases, we used Microsoft's DREAD Risk Assessment Model. \footnote{http://msdn.microsoft.com/en-us/library/ff648644.aspx} This model assigns values to different factors influencing threats. The various factors which are used to rank a threat are:
\newline
1. Damage: How big the damage would be if the attack succeeded?
\newline
2. Reproducibility: How easy it is to reproduce the attack to work?
\newline
3. Exploitability: How much time, effort, and expertise is needed to exploit the threat?
\newline
4. Affected Users: If a threat were exploited, what percentage of users would be affected?
\newline
5. Discoverability: How easy is it for an attacker to discover this threat?
\newline
\newline
A simple scheme such as High (3), Medium (2), and Low (1) is used to rate the system to calculate the risk value associated with it.
Referring to the information portals maintained by the government of Delhi, it is possible to assign values to the DREAD factors as:

\textbf{Threat: Malicious user can identify PII of Delhi residents.}
\begin{itemize}
  \item[$\bullet$ ] \textbf{Damage Potential:}
  Threat to reputation of a person and legal liability because of leaking sensitive information: 2
  \item[$\bullet$ ] \textbf{Reproducibility:}
  Fully reproducible: 3
  \item[$\bullet$ ] \textbf{Exploitability:}
  A skilled programmer can launch the attack and repeat the steps: 2
  \item[$\bullet$ ] \textbf{Affected Users:}
  All users, default configuration: 3
  \item[$\bullet$ ] \textbf{Discoverability:}
  Published information explains the attack: 3
\end{itemize}

\begin{figure*}[!htp]
\centering
%\vspace{-2.0cm}
\includegraphics[scale=0.40, angle=0]{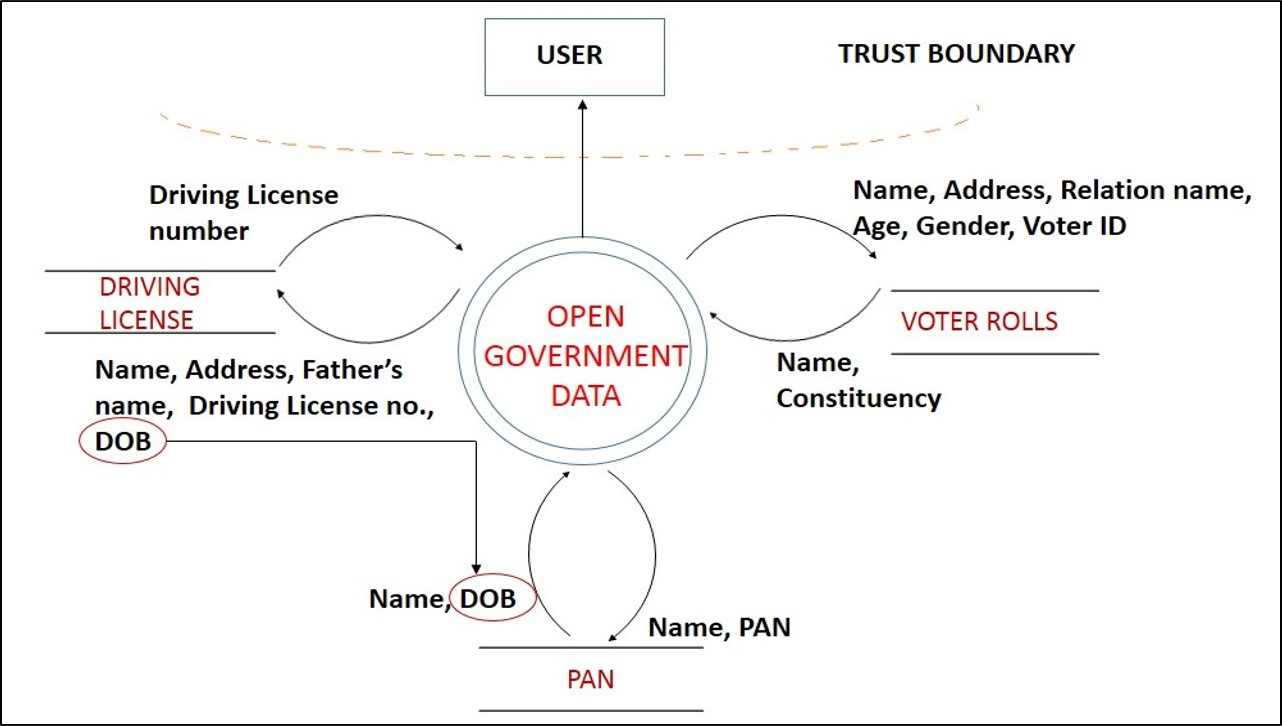}
\vspace{0.5cm}
\caption{Data Flow Diagram (DFD) for various open government databases.}
\label{fig:threat}
\end{figure*}

The overall rating is 2 + 3 + 2 + 3 + 3 = 13 which is considered high. We can thus conclude that this threat pose a significant risk to the various information portal websites of Delhi government and needs to be addressed as soon as possible.
\newline
Data Flow Diagrams (DFD) can be used to graphically represent a system. Trust boundaries are added to show the threat modelling. Figure \ref{fig:threat} shows the Data Flow Diagram (DFD) depicting the data flow between a user and the various open government data sources. The trust boundary shows the border between the trusted and untrusted elements. An attacker, being an untrusted element can get the details of another user by exploiting the vulnerabilities of open government sources.
\section{Data Extraction}
The various open government data sources mentioned above do not provide APIs for data access. The only way to access the data is through the web interface. This requires HTML-parsing and page-scraping techniques which make the system little prone to errors. The seed information and the result set for each source was different to be discussed separately.
\subsection{Driving Licence Number}
Since the Driving Licence number exist in sequential order, we could collect a large volume of data. We took 5 different driving licence numbers belonging to different students at the university. Each of these five driving licence numbers are then sequentially incremented (upto 300) and queried against the database and the results are stored in locally maintained data bank for later retrieval and analysis.
\subsection{Voter ID Number}
The mandatory inputs to Voter ID government portal is the constituency number and name of the person. As stated earlier, there are 70 constituencies in Delhi region. Current state of the art gives results of all the individuals starting with a particular alphabet when the input to the system is a single alphabet, say `a'. The result pool is looped for all the 70 constituencies and stored in the local database.
\subsection{Permanent Account Number (PAN)}
The mandatory inputs to the PAN card income tax portal is the name (last name, first name) and the DOB. This information was obtained from the Driver's license database. It was then used to query the PAN number portal and the results were stored in a separate database for analysis.
\subsection{Mahanagar Telephone Nigam Limited (MTNL) Phone Numbers}
The only input the MTNL phone directory portal was the name of the person. However, with the current state of the art, just entering a random character (for e.g., say `a'), it gave information about all the individuals whose name start with `a'. The portal is queried for all the 26 English alphabets.
Figure \ref{fig:govt} below shows the architecture diagram for OCEAN to extract data from open government data sources.
\section{Information Aggregation}
The final step was to combine the information that was obtained from various sources. The database described above gave some unique information about the user. The aggregation of information was important because a user might hide some personal attributes from a particular social network, but after aggregation we can get comparatively large amount of information. E.g., the location attribute is not available publicly through Facebook, but when that user is found to be connected to other networks; his location could be derived from the latter. The aggregation could be performed on three networks i.e., Twitter, Foursquare and Facebook. This was possible since some of the foursquare users mentioned Facebook and / or Twitter accounts on their foursquare profile. PHP scripts were written to get data from the government websites. The data collected from data sources were stored in the MySQL tables. Since a user can exist in both the databases, we tried to aggregate such users together. This was done on the basis of name, address of the user. There is a lack of consistency in the various terminologies and methodologies employed by different authorities. Therefore, this aggregation was challenging since the format of address storage differs across the databases and hence decreased the number of users that could be aggregated. The aggregated results were visualized using a Family tree which shows information about his / her parents, siblings and spouse.
\begin{figure*}[!htp]
\centering
%\vspace{-2.0cm}
\includegraphics[scale=0.45, angle=0]{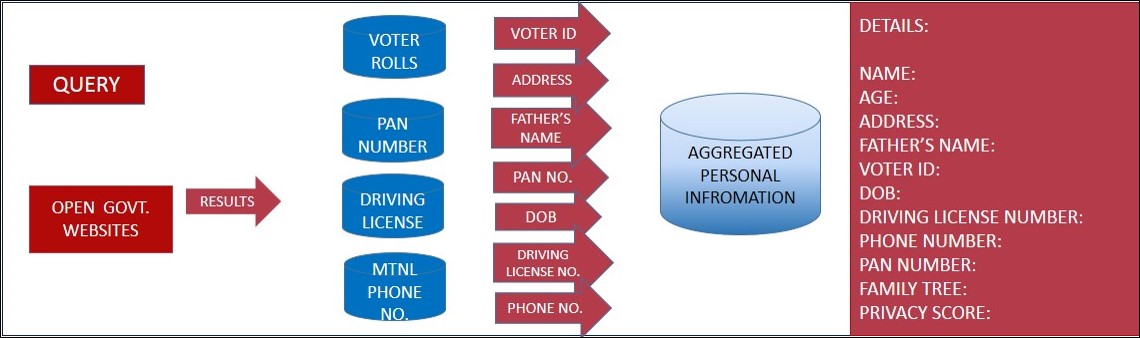}
\caption{Architecture diagram for obtaining data from open government data sources.}\label{fig:govt}
\vspace{.5cm}
\end{figure*}
\subsection{Social Networks}
Public API calls were used to collect information from social networks. The names which exist in the government databases were used to query the OSN’s. Twitter's REST API v1 and v1.1 were used to get the data. OAuth tokens were used for authorization and authentication. Facebook Graph API gave access token for getting the data from Facebook servers. Foursquare API v2 was used to search the users from foursquare. Calls through Google Plus was made with authenticated keys unique to the user account. LinkedIn used OAuth tokens to authorize data collection. The process of data collection was slow and time demanding since API calls that are used to extract data are rate limited. For e.g., the daily limit for GooglePlus to extract the user information
is 10k requests/day and the number of profiles on GooglePlus is high, the rate limit exceeded
very frequently and hence slowed down the process. Figure \ref{fig:social} below shows the architecture diagram for OCEAN to extract data from various online social networks.
\begin{figure*}[h!]
\centering
%\vspace{-2.0cm}
\includegraphics[scale=0.35, angle=0]{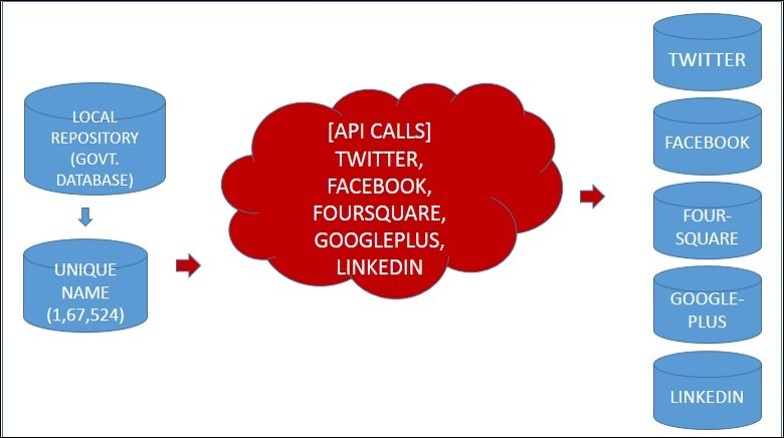}
\caption{Architecture diagram for obtaining data from social networks.}\label{fig:social}
%\vspace{.5cm}
\end{figure*}

\chapter{System Design}
\begin{figure*}[!htp]
\centering
\vspace{0.5cm}
\includegraphics[scale=0.40, angle=0]{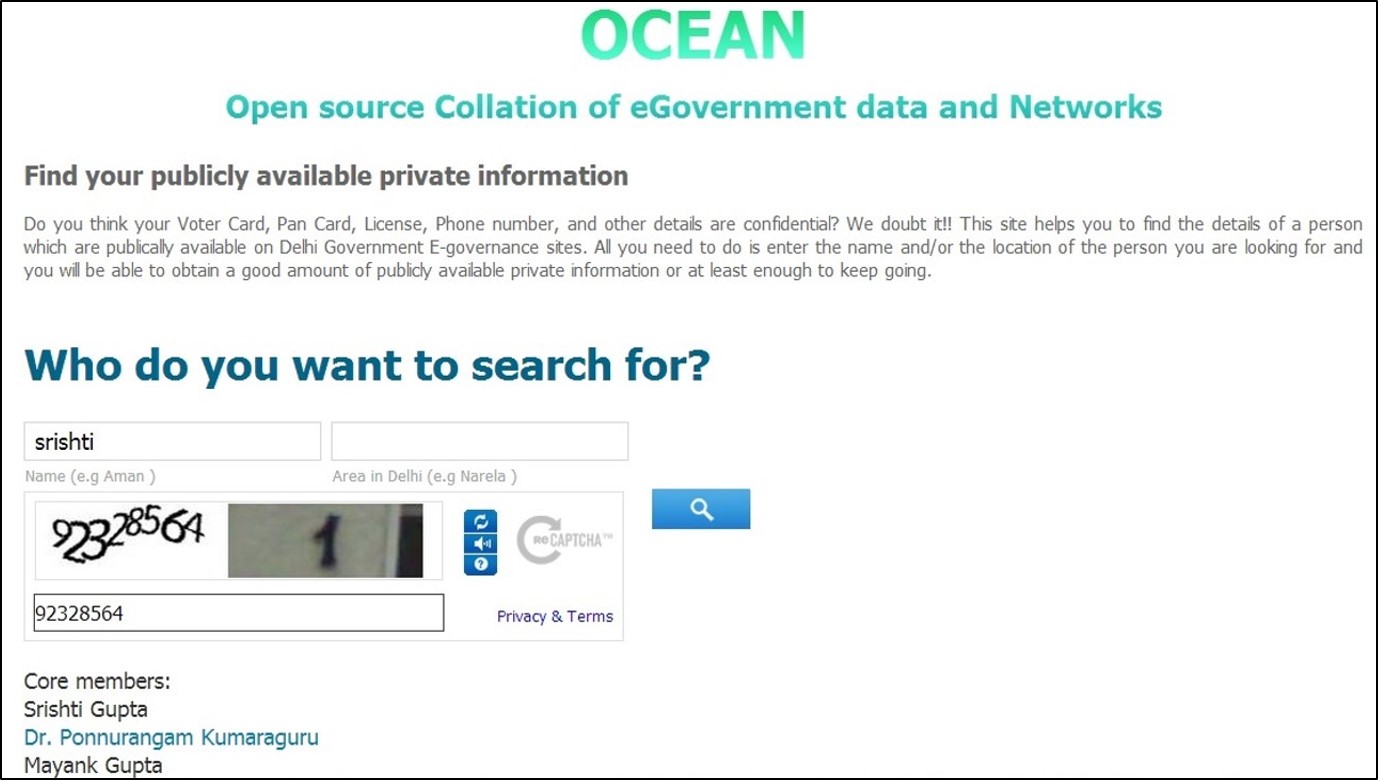}
\vspace{0.5cm}
\caption{Screen 1 of the system where the user is required to enter the name, location (optional) of the person he wants to search and a CAPTCHA to verify himself as a human.}
\label{fig:ocean-1}
\end{figure*}

\begin{figure*}[!htp]
\centering
\includegraphics[scale=0.45, angle=0]{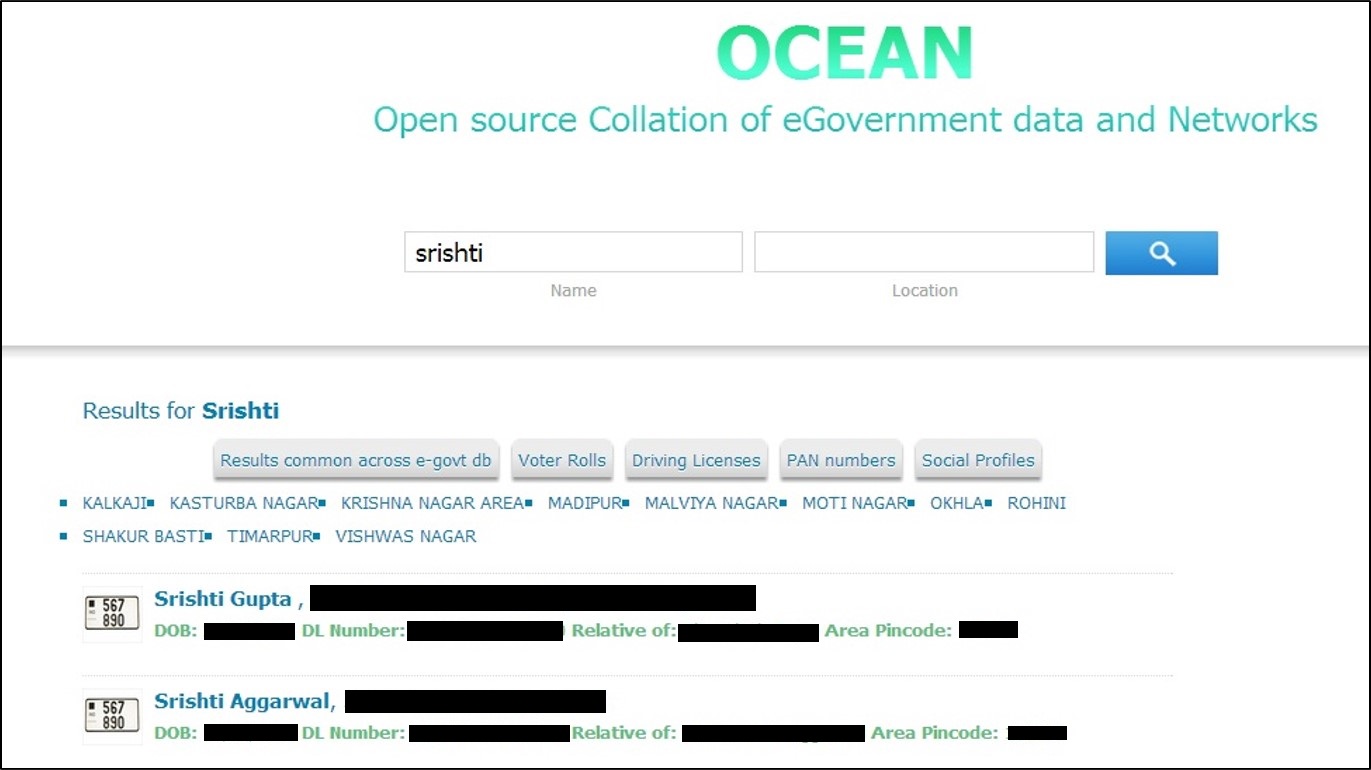}
\vspace{0.5cm}
\caption{Results for only the driving licence database based on the user query.}
\label{fig:ocean-3}
\end{figure*}

\begin{figure*}[!htp]
\centering
\includegraphics[scale=0.45, angle=0]{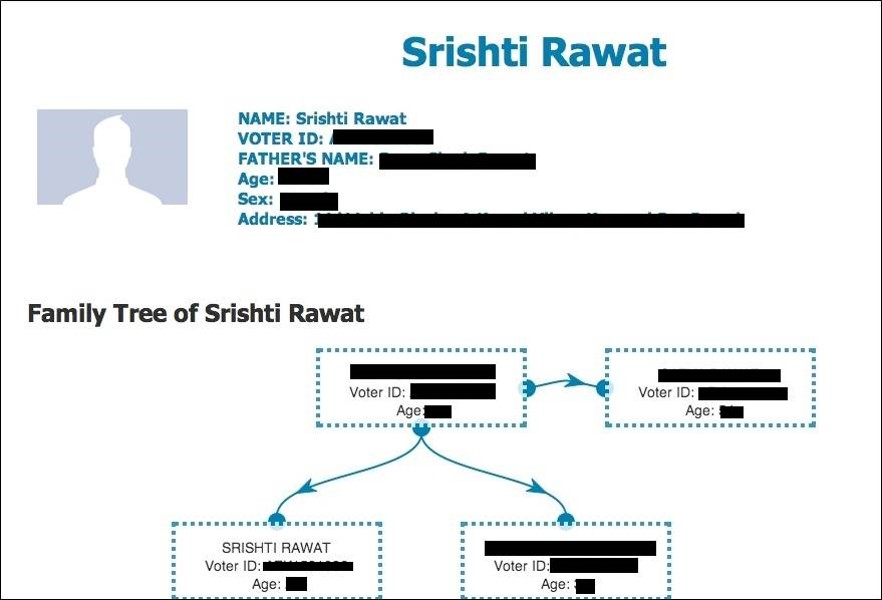}
\vspace{0.5cm}
\caption{Family Tree of the user, created by information aggregation shows the details of parents and siblings of the user.}
\label{fig:ocean-4}
\end{figure*}

\begin{figure*}[!htp]
\centering
\includegraphics[scale=0.45, angle=0]{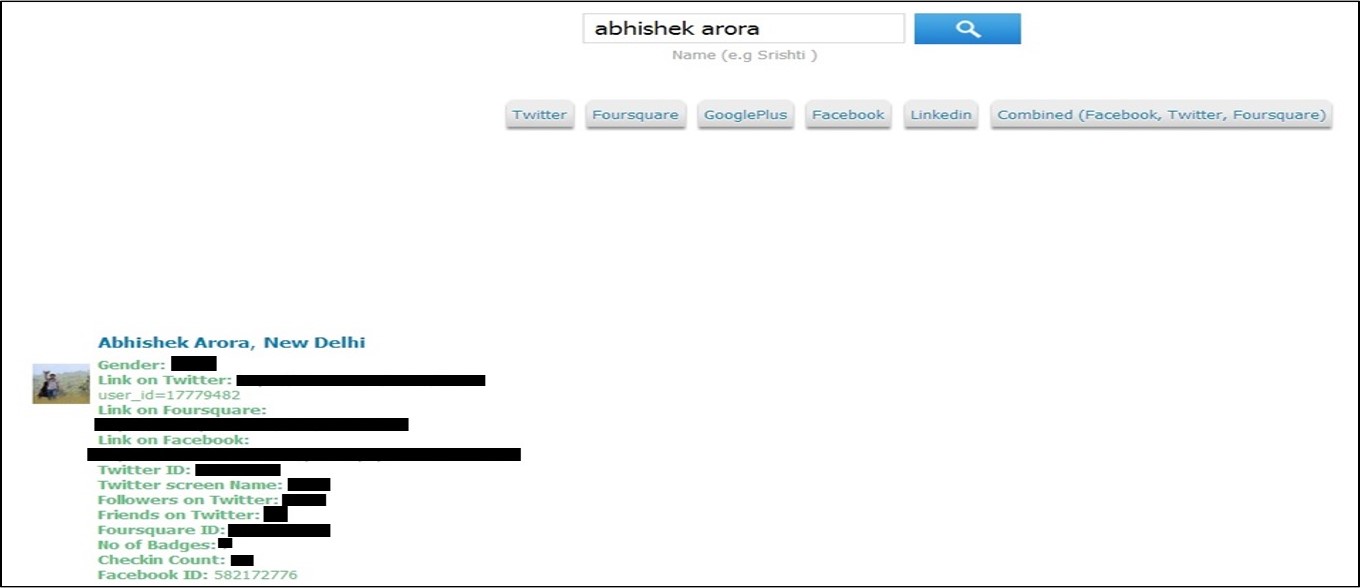}
\vspace{0.5cm}
\caption{Combined details of the user from Facebook, Twitter and Foursquare.}
\label{fig:map-all}
\end{figure*}

Our work is divided into two parts. In the first part, we did a threat modelling on various open government data sources to understand the risks and threats arising due to the public availability of personal information of the residents of Delhi from these repositories. We then created a system called OCEAN: Open-source Collation of eGovernment data And Networks which brings together the information from these sources under one roof. The other part was a user study to measure the accuracy, effectiveness and usability of the system. The study also focussed on understanding the perceptions, reactions and expectations of the people after using this system.
In this section, we elaborate the results from the first part of our work, i.e. results obtained after building OCEAN.
\newline
OCEAN was made available to the general public on January 21, 2013. Since then, it has recorded 407 \emph{unique visitors} (as on October 10, 2013). OCEAN has 8,195,053 Voterrolls; 2,24,982 Driving licences; 53,419 PAN card numbers; 1,557,715 Twitter; 3,377,102 Facebook; 29,393 Foursquare; 1,86,798 LinkedIn and 28,900 GooglePlus records.
Figure \ref{fig:ocean-1} shows the first screen of the system where the user enters the name and location (\emph{location}) of the person he wants to search. The user is also required to enter a CAPTCHA which is implemented as a security feature for the system.
The user is then re-directed to the result page where he can view the results based on his query. The page shows the results for Voter ID, driving licence number, PAN numbers and the social networks as shown in the Figure \ref{fig:ocean-1}. The page also shows the results of the users existing across the various open government sources under the \emph{`Results across the e-govt db'} tab.
\begin{figure*}[!htp]
\centering
\includegraphics[scale=0.45, angle=0]{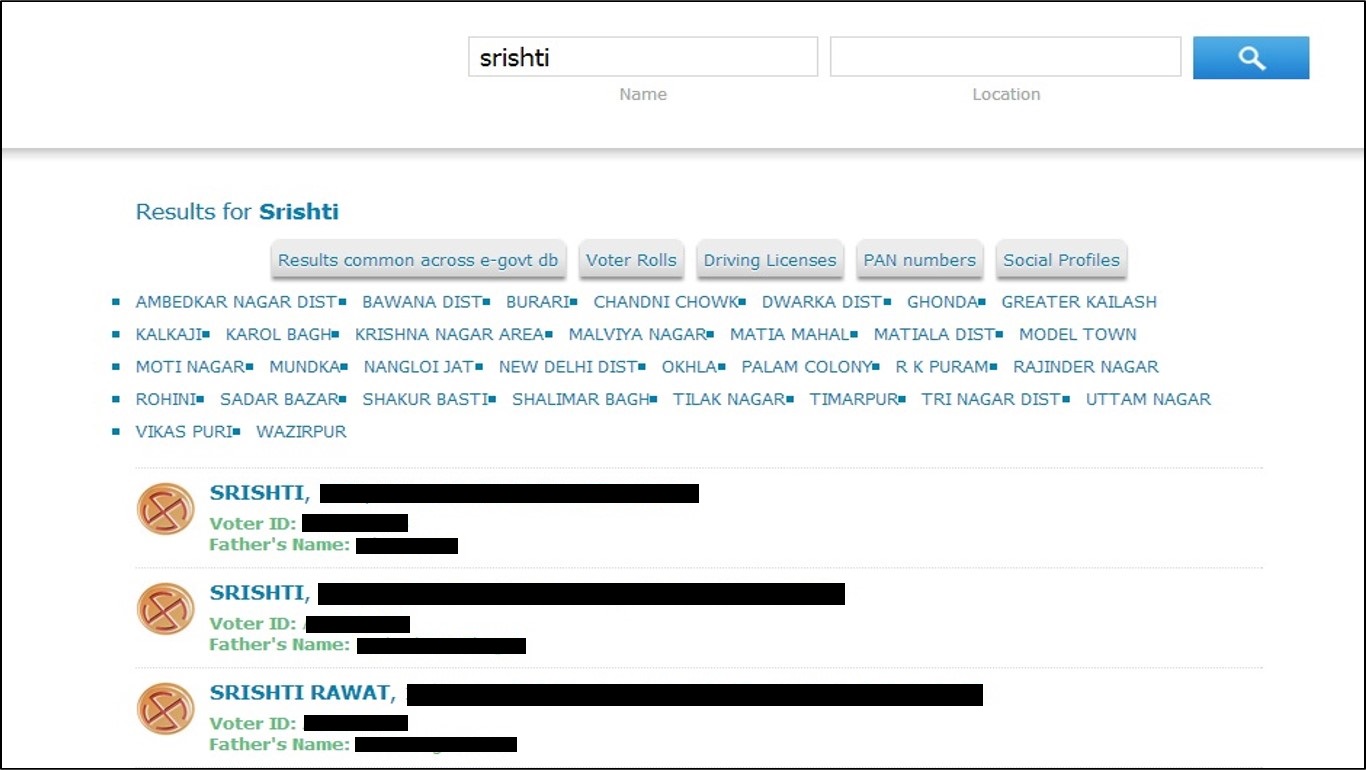}
\vspace{0.5cm}
\caption{Screen 2 of the system where the user is shown the results from the system for his query.}
\label{fig:ocean-2}
\end{figure*}
The system results for specific open government database can be viewed by clicking the respective tabs. Figure \ref{fig:ocean-3} shows the result-set for driving licence number database of Government of Delhi.
Figure \ref{fig:ocean-4} shows the \emph{family tree} of the person which is obtained by aggregating the information within the Voter ID database. This tree pictorially represents the details of parents and siblings of the user. OCEAN has 3,03,393 users whose family tree can be displayed.
\newline
Apart from the open government databases, OCEAN also gives the personal information about a user from social networking sites like Facebook, Twitter, GooglePlus, LinkedIn and Foursquare. The tab \emph{Combined} shows the results of the user on aggregation from Foursquare, Facebook and Twitter. Foursquare allows user to specify their Facebook and Twitter account. For such set of users, the three databases are combined and displayed. OCEAN has 11 such users. Figure \ref{fig:map-all} shows the results obtained for one such random user whose details are available from Facebook, Twitter and Foursquare.
\newline
\newline
OCEAN is scalable in obtaining data from other information sources like criminal records of a person, electricity bill payment etc. The data can also be collected from other states in India.
\chapter{Experiments and Analysis}
In this section, we will focus ourselves to the analysis of the data obtained from the survey dataset. This section talks about calculating privacy score of the people existing in OCEAN, accuracy and system usability score of the system. This section also gives an empirical analysis about the reactions, perceptions and expectations of the people after using the system.
\section{Survey Dataset}
We collected data from an online survey which was distributed to the university's mailing list and shared on the social networking sites. People were asked to use the system and fill the survey study. We have received 62 complete responses. Among the participants, 51\% are males and 49\% are females, 77\% are in the age of 20 to 25, 60\% hold a college degree or higher. 23\% participants had either self or friends experienced with identity thefts online.
\section{Privacy Score}
A privacy score of the user measures the risk associated with the person on the leakage of his / her personally identifiable information from the open government databases. The user is at risk if this information is available freely to everyone since it can be used with malicious intents. Harel et.al \cite{harel2010m} calculated a similar M-score: Estimating the Potential Damage of Data Leakage
Incident by Assigning Misuseability Weight, of the relational databases maintained by the organizations to handle employee data. It estimates the damage which can result if organization's data is exposed and falls into wrong hands.
\newline
\newline
Sensitivity score function: The sensitivity score function f: C * S$_j$ $\rightarrow$ [0,1] assigns a sensitivity score to each possible value x of S$_j$, according to the specific context c $\epsilon$ C in which the table was exposed. The sensitivity score is defined by the user as it reflects the data owner's perception of the data's
importance in different contexts. People will consider the information as private / public in different contexts. For e.g., giving the bank account balance when talking with mortgage consultant against a stranger on the street.
For each record r, x$_r$ of S$_j$ as S$_j$[x$_r$].
\newline
RRS$_i$: Raw Record Score for record i, calculated from the summation of the sensitivity score of all the attributes in that record and
\newline
D$_i$: Distinguishability factor which determines the uniqueness of the record.
\newline
x: Settable parameter for adjusting the effect of the table size on the final score.

\begin{algorithm}[!htp]
\small \caption{Algorithm for calculating M-score.}
\label{Algo}
\SetAlgoLined
\small \KwIn{A set of `r' records for users in the organization.}
\small \KwResult{M-score of the user.}
\small initialization\;
\small Assign sensitivity score to each record.
\newline \small \For {each record i in R} {
\begin{eqnarray}
RRS_{i}=\min(1, \sum_{S_j\epsilon T} f(c, S_j[x_i]))
\end{eqnarray}
\begin{eqnarray}
Mscore&=&r^{1/x}*RS=r^{1/x}*\max_{0 \leq i \leq r}( \frac{RRS_{i}}{D_i})
\end{eqnarray}
\eIf{ (Sensitivity of data insignificant)} {
\small x = 1\;
}
{
x = $\infty$;
}
}
\end{algorithm}
We now show how we can sue this equation in our case of open government databases:
\subsection{Number of records, `r' in all the open government databases}
The number of records / users, `r' specific to certain open government databases with the corresponding attributes is shown in the Table 5.1. For e.g., driving licence contains the user whose entry is present only in the driving licence database, not in Voter ID and pancard.
\begin{table*}[!htp]
\caption{Number and attributes for each open government database.}\label{tab:metadata}
\begin{center}
\begin{tabular}{|p{3cm}|p{3cm}|p{9cm}|} \hline
\textbf{Database} & \textbf{Number of Records} & \textbf{Attributes} \\ \hline
\small Driving licence & \small 1,65,121 & \small DL number, Name, Address, Relative's name, DOB.\\ \hline
\small Voter ID & \small 81,88,669 & \small Voter ID, Name, Address, Father's / Mother's / Husband's name, Age, Gender.\\ \hline
\small PAN card & \small 53,419 & \small PAN Number, Name, DL number, Address, Relative's name, DOB.\\ \hline
\small Voter ID + driving licence & \small 6,384 & \small Voter ID, DL number, Name, Address, Father's / Mother's / Husband's name, Age, Gender.\\ \hline
\small Voter ID + driving licence + PAN number & \small 1,693 & \small Voter ID, DL number, PAN number, Name, Address, Father's / Mother's / Husband's name, Age, Gender.\\ \hline
\end{tabular}
\end{center}
\end{table*}
\subsection{Sensitivity Score}
The sensitivity score relies on the perceptions and attitude of the people as calculated from our survey study. One of the goals of the survey was to collect user's information-sharing preferences. Given a list of proﬁle items that span a large spectrum of one’s personal life (e.g., Name, Gender, DOB, Address, Voter ID, driving licence number, PAN number, Age, Father's name), the users were asked to specify whether they want to share this information online. Since in our case, the government databases are visible to everyone, we only consider the case where the user does not want to put his / her information online / not share with everyone. According to the proportion of users marking it non-sharable, a privacy level is allocated values \{1,2,3,4,5\} to each attribute where level `1' means people are not very restrictive in sharing their information with everyone and `5' meaning people do not want to share their information with anybody. Thus, the sensitivity score for each attribute is directly mapped with the privacy level for each attribute. Higher the privacy level, more sensitive is the information. Table 5.2 below shows the response of the users from the survey for the attributes relevant in our case.
\begin{table*}[!htp]
\caption{Response of the survey users showing their unwillingness to share their personal information with everybody.}\label{tab:metadata}
\begin{center}
\begin{tabular}{|p{3cm}|p{4cm}|p{3cm}|} \hline
\textbf{Attribute} & \textbf{Percentage of users unwilling to share with anybody} & \textbf{Privacy level} \\ \hline
\small Voter ID & \small 56.4\% & \small 4\\ \hline
\small Driving licence number & \small 58\% & \small 4\\ \hline
\small PAN & \small 67.7\% & \small 5\\ \hline
\small Full name & \small 14.5\% & \small 1\\ \hline
\small Home Address & \small 82.2\% & \small 5\\ \hline
\small Age & \small 29\% & \small 2\\ \hline
\small DOB & \small 50\% & \small 3\\ \hline
\small Father's name & \small 38.7\% & \small 3\\ \hline
\small Gender & \small 14.5\% & \small 1\\ \hline
\end{tabular}
\end{center}
\end{table*}
\subsection{Distinguishability Factor and `x'}
For our scenario, we take the value of x = $\infty$ since the sensitiveness of the information is the key for our evaluation. Also, the distinguishability factor, D$_i$ of our attribute set is 1 since every user has a unique key to uniquely identify itself (Voter ID, driving licence number, PAN number).
\newline
\newline
Our equation is thus reduced to
\begin{eqnarray}
Privacy Score=r^0*\max_{0 \leq i \leq r}(\frac{RRS_{i}}{1})
\end{eqnarray}
or
\begin{eqnarray}
Privacy Score=\max_{0 \leq i \leq r}RRS_i
\end{eqnarray}
where
\begin{eqnarray}
RRS_{i}=\sum_{S_j\epsilon T} S_j[x_i]
\end{eqnarray}
Using this equation, the Privacy Score (PS) for all the users in each of the open government databases can be calculated as:
\begin{itemize}
 \item[$\bullet$ ] \textbf{Case 1: Users having only Voter ID (97.3\%)}
 \newline
 PS = $\sum{\text{(Voter ID, name, father's name, age, gender, address)}}$ = 16
 \item[$\bullet$ ] \textbf{Case 2: Users having only driving licence number (2\%)}
 \newline
 PS = $\sum{\text{(DL number, name, relative's name, DOB, address)}}$ = 17
 \item[$\bullet$ ] \textbf{Case 3: Users having only PAN number (1\%)}
  \newline
 PS = $\sum{\text{(PAN number, DL number, name, relative's name, DOB, address)}}$ = 25
 \item[$\bullet$ ] \textbf{Case 4: Users having Voter ID and DL number (0.07\%)}
  \newline
 PS = $\sum{\text{(Voter ID, DL number, name, father's name, age, gender, DOB, address)}}$ = 24
 \item[$\bullet$ ] \textbf{Case 5: Users having Voter ID, DL number and PAN number (0.02\%)}
  \newline
 PS = $\sum{\text{(Voter ID, DL number, PAN number, name, father's name, age, gender, DOB
 , address)}}$ = 29
\end{itemize}
The person with higher privacy score is at greater risk since more number of personal attributes are available to uniquely identify him / her online. 1,693 people, whose privacy score is 29 are at maximum risk and most vulnerable to attacks. We have multiple personal attributes which can uniquely identify that person like Voter ID, Driving license number, PAN number, home address. It then becomes easy to perform targeted personal attacks against these people. For e.g., getting fake Voter ID cards, fake bank accounts, multiple SIM cards issued in their name etc. As discussed in chapter 1, is also possible to register them on some vulnerable sites (for e.g., Income tax payment portal) and view highly sensitive information like the TDS statement.

\section{Evaluation Metrics}
\subsection{Recall and System Usability Score}
The system recall is defined as
\begin{eqnarray}
Recall=\frac{\text{Number of people who could be identified in the system}}{\text{Total number of search operations done on the system}}
\end{eqnarray}
Thus, Recall = (179 / 389) * 100 = 46\%.
\newline
The recall is low since web collection does not give 100\% results. We have around 8 million of the total 12 million Voter ID records \footnote{http://m.indianexpress.com/news/electoral-roll-complete-delhi-has-1.23-cr-voters/1057665/}. As part of our future work, we plan to expand our database.
\newline
\newline
The System Usability score (SUS) is measured using the standard method by Brooke et.al.\cite{brooke1996sus} For OCEAN, the values of SUS comes out to be 74.5 / 100 which means that people found the system usable and convenient to use.
\section{User Experience and Expectations}
\begin{figure*}[!htp]
\centering
\includegraphics[scale=0.45, angle=0]{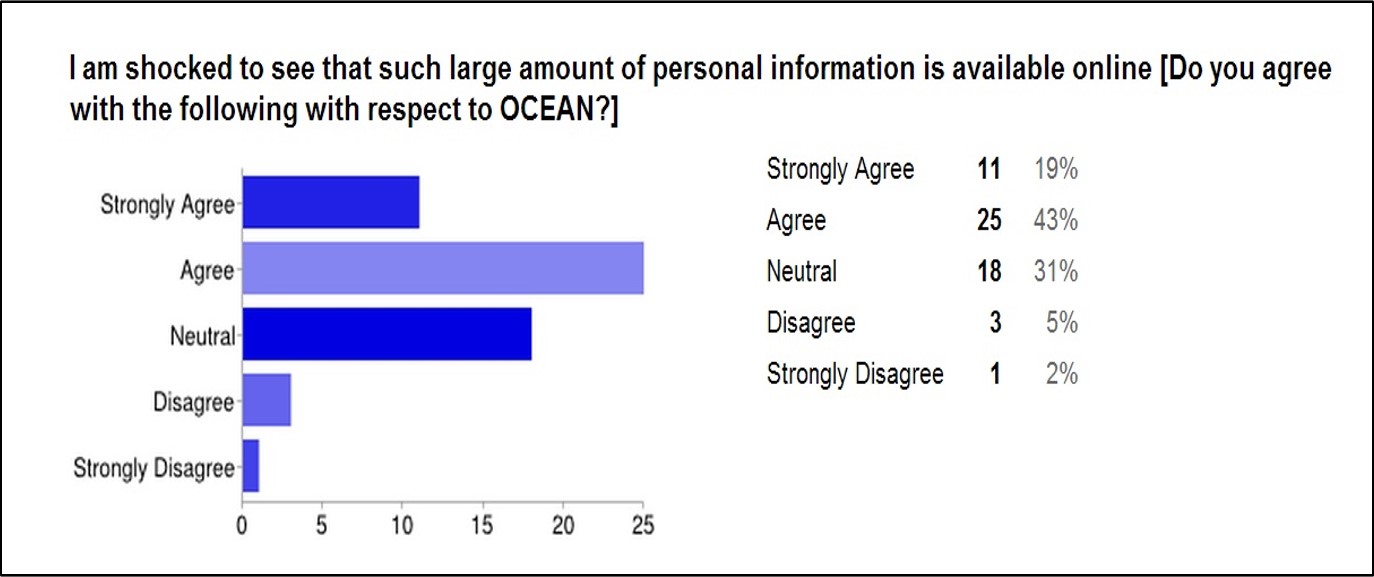}
\caption{Screenshot showing how shocked users were after using OCEAN.}\label{fig:shocked}
\end{figure*}

\begin{figure*}[!htp]
\centering
\includegraphics[scale=0.45, angle=0]{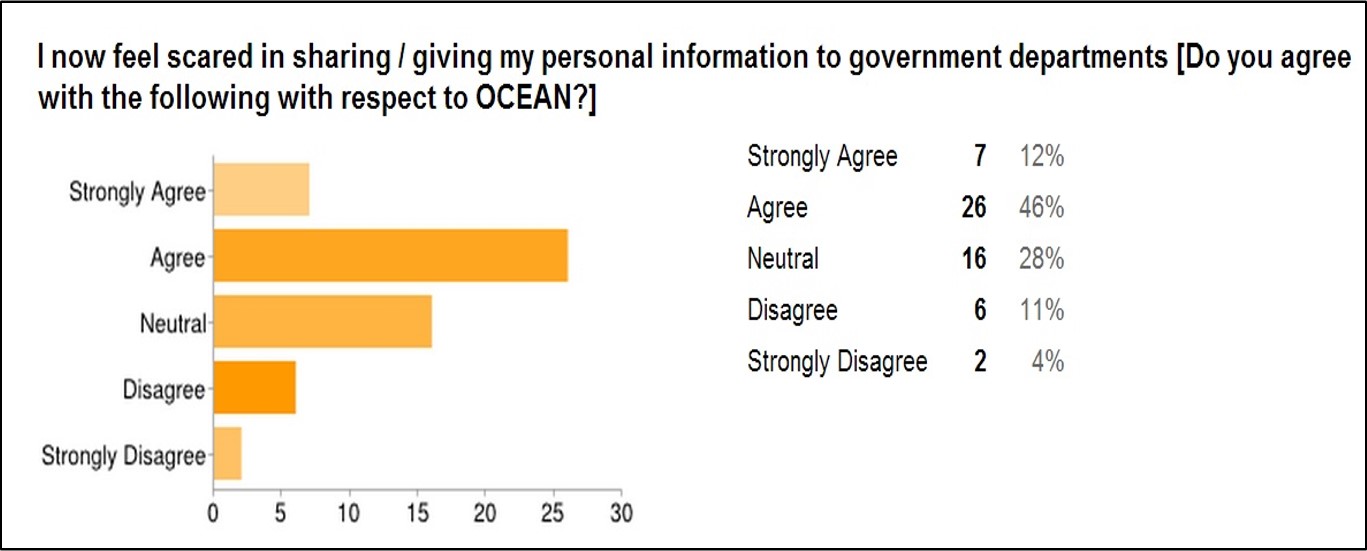}
\caption{Screenshot showing how much users were scared after using OCEAN.}
\label{fig:scared}
\end{figure*}

In this section, we talk about the privacy awareness of the citizens about the open government data, their reactions to the availability of this information after using OCEAN and expectations from the government departments handling the public databases in how to protect this information.
\newline
Ironically, even though the government has started various open government initiatives to increase the level of transparency with it's citizens in terms of their data, majority of the people are still unaware of these services. According to our survey results, only 19\% people are aware of the existing online public databases like Voter ID, PAN number, Electricity bill payment, Income tax payment and driving licence portals. Around 76\% have only started using these services for less than 2 years. This shows that although the government has given access to the data in digital form for easy delivery of government services to it's citizens, it is not being properly conveyed to the general public. Hence, the government needs to come up with proper scheme and plans to convey the existence and utility of these systems to the public effectively.
\newline
\newline
However, once the citizens are aware of the existence of these open government databases after using OCEAN, they are not really comfortable with their personal information put online. The reactions of users are captured in Figure \ref{fig:shocked} and Figure \ref{fig:scared}. According to our study, 62\% of the respondents were shocked to see that such large amount of personal information is accessible to everyone and said that they now feel reluctant in sharing their information with the various government departments. When the participants were asked (Question: ``The availability of this personal information does not bother me as it does not harm me personally"), 57\% were against this statement and felt that this information could be used maliciously against them. Some of the feedback for the system, which shows the sentiment of the users are, \emph{`The system's implications and functionality are undoubtedly great. I am really shocked that the exact ID numbers are available online without much security against data mining at this scale.', `It was an eye-opener to a common man and arouse the curiosity to disclose personal details very cautiously.'}.

Given that the people are now aware of the large amount of information available online by several government departments, we now discuss the expectations of people from these government authorities.
43\% people thought that there is no need to put the personal information about the individuals online. However, 52\% felt that the information could be available online but with proper rules and regulations so that the accessibility of the information to everybody could be restricted. When asked about what personal information should not be put online, around 65\% of the participants felt that uniquely identifiable information like Voter ID, driving licence number, PAN number, home address, MTNL phone number should not be made available. 8 - 9\% people said that details like name, age, gender should not be put online while 15\% people felt uncomfortable sharing their DOB and father's name to everyone online. In response to the expectations of the people from the government on the availability of such large amount of personal information online, 43\% people felt that authorization mechanisms (username, password) should be implemented for accessing the data. This helps in a way that it will allow a user to view only his / her details rather having access to everyone's information. 21\% of the participants were in favour of having security measures like CAPTCHA which prevents data collection on the government websites which give access to these databases. Searching for a large set of users manually is little cumbersome than writing scripts to extract all the data at once. Solutions like CAPTCHA restrict the implementation of these scripts to some extent. This could be a first step in protecting the information by government of Delhi on it's public servers. 34\% people felt that proper privacy laws should be created so that even though the government needs to put this information for streamlining its own information gathering and processing procedures, legal liabilities must exist against the hackers and malicious users. It is interesting to see that only 2\% of the participants felt that the data should be removed immediately. This shows that people want easy access to their information provided that proper security measures and privacy laws are made to govern the accessibility of such personal information.
\chapter{Conclusion, Limitation and Future Work}
In this chapter, we will cover the conclusion of our work, discuss the limitations and scope of improvement as the future work.
\section{Conclusion}
On developing OCEAN:Open-source Collation of eGovernment data And Networks, we demonstrated that large amount of authentic data is available on government servers like Voter ID, driving licence number, PAN number. Querying this database with little seed data gives a lot of personal data about the target. The ease with which this data can be extracted calls for addition of security features to be put on such sensitive data. The fact that the source of data is issuing authority (Government of India departments) itself makes it even more valuable as it comes with guarantee of being mostly accurate.
\newline
Given the range and volume of data that we were able to collect during the course of our thesis work, it is easy to conclude the increasing privacy threats to any individual covered in the databases mentioned. In fact, in some cases the query parameters can be sequentially generated, making it easier to gather data in bulk. With the interlinking (output data from one source as input to the
other) of data-sources, we can achieve the horizontal-depth for every subject in the database. It is for this reason that, identification of any new such sources adds both the horizontal and vertical (volume) of the collected data. The PII collected apart from being a serious privacy violation, can be used to draft personalize attacks against individual including impersonation (Many scenarios use some of this information like mothers maiden name, PAN as additional information to confirm an individuals identity).
\newline
A threat modelling on these open government databases and a DREAD score of `13' shows the level of risk from these data sources and ascertains that the government needs to ensure that the data available on its servers is not used by people with malicious intent.
\newline
An privacy score in the range 16 - 29 for the people existing in OCEAN suggests that the Delhi residents are at risk and their privacy is at stake due to the availability of their PII online. The recall of the OCEAN was found to be 46\% and SUS score of 74.5 which shows that the users of OCEAN found it usable and convenient to use.
\newline
A survey study of 62 users showed that people were shocked to see that such large amount of information exists on the government databases. It also showed that people feel that this information could be used maliciously against them. Majority of the people felt that the government should implement authorization mechanisms like username and password on the websites revealing this information to restrict the information availability to anyone and everyone.
\newline
\section{Limitation}
Datasets available are limited to the residents of Delhi. Since the data is collected from the government websites, the dataset is not 100\% and hence the recall is less. For creating the family tree, the record for the person and his father are matched, spelling mistakes of the same person in multiple government database sources put a constraint in aggregation. We have limited users across multiple government databases. Since the address format for Voter ID and driving licence was different, it was difficult to map the users across both the databases.
\section{Future Work}
As a part of future work, the database can be expanded which was limited since complete data could not be collected from the government websites. This will help to improve the recall of the system. One can also incorporate national level databases i.e., databases existing in other states of India into OCEAN and show the privacy leakage, if any, for that state. One can also look into the UID Aadhar card database to see if any privacy breach exist and general public can get access to the complete database.
\newline
Instead of broadly looking at all the Delhi citizens, one can specifically focus on a single user by collecting his information from all possible sources on the web. It can be used to calculate a reputation score for the person in the society. For e.g., collecting information from the criminal records of a state can be used to calculate such score. If a person is found in that record, it will assign a negative score to the person whereas a person who cannot be found in any of such records will get a better reputation score in the society. An aggregation of all this information can help compare a person`s image in the society.
\newline
One can also aggregate results from social media to find attributes about an entity and hence map that user with the data obtained from open government data sources. This helps in connecting an offline identity (as in govt. sources) with an online identity (as maintained on online social networks) of a user. Since the public attributes available from the social media are limited, one needs to adopt a proper methodology to extract maximum amount of information from these social networking sites.

%\chapter*{Appendix}\label{chapter:appendix}

\end{document}